\documentclass[preprint,3p,times]{elsarticle}
\usepackage{amssymb}
\usepackage{amsmath}
 \usepackage{lineno}
 \usepackage{natbib}
 \usepackage{xcolor}

\journal{TBD}

\begin{document}
\begin{frontmatter}

\title{Fracture of disordered and stochastic lattice materials}

\author[label1]{Sage Fulco \corref{cor1}}
\author[label1]{Prashant K. Purohit}
\author[label2]{Michal K. Budzik}
\author[label1]{Kevin T. Turner \corref{cor2}}

\affiliation[label1]{
            organization={Department of Mechanical Engineering and Applied Mechanics, University of Pennsylvania},
            city={Philadelphia},
            state={PA},
            country={USA}}
\affiliation[label2]{
            organization={Department of Mechanical Engineering and Production, Aarhus University},
            city={Aarhus},
            country={Denmark}}
            
\cortext[cor1]{Corresponding author at: Department of Mechanical Engineering and Applied Mechanics, University of Pennsylvania, Philadelphia, USA. E-mail address: fulco@seas.upenn.edu (S. Fulco)}
\cortext[cor2]{Corresponding author at: Department of Mechanical Engineering and Applied Mechanics, University of Pennsylvania, Philadelphia, USA. E-mail address: kturner@seas.upenn.edu (K.T. Turner)}

\begin{abstract}
The failure of mechanical metamaterials is a function of the interplay between the properties of the base material and the microstructural geometry. Stochastic failure properties of the base material and disordered microstructural geometries can contribute to variations in the global failure mechanics that are not captured in traditional analyses of ordered, deterministic architected materials. We present a probabilistic framework that couples stochastic material failure and geometric disorder to predict failure in lattice mechanical metamaterials. These predictions are verified through finite element analysis, which confirm that disorder and stochasticity affect both the mean and variance of the damage initiation load in a lattice, with average failure loads being generally reduced and variance increasing with higher levels of disorder and stochasticity. The fracto-cohesive length and representative volume element size are also predicted and constrain the minimum defect and lattice sizes, respectively, for failure to be considered a fracture process.  The framework is extended to consider the fracture behavior of the lattice, the development of damage zones, and their impact on the steady-state fracture toughness.
\end{abstract}




\end{frontmatter}


\section{Introduction}
\label{sec:intro}
The fracture properties of many materials emerge from their structural features at varying length scales. Fundamentally, the localized nature of fracture leads to the nucleation and spread of damage throughout fine-scale microstructures before ultimately resulting in global failure of the material. These microstructures, whether naturally occurring in high toughness materials like bone \cite{Tertuliano_NatMat_2016,Yang_Biomat_2006} or in low toughness materials such as blood clots \cite{Garyfallogiannis_ActaBio_2023, Purohit_JMPS_2025}, play a critical role in how damage develops near a crack tip, which in turn impacts the material's toughness. In traditional engineering materials, knowledge of material microstructure can be used \cite{Bazant_IntJourFrac_1990,Bazant_JourEngMech_1990} to inform predictions of fracture toughness; however, these structures, whether regular or random, are often inherent to the material or controlled rudimentarily through manufacturing process controls. Conversely, in structural mechanical metamaterials, precisely engineered fine-scale structures are designed and leveraged to control and enhance failure properties \cite{Hossain_JMPS_2014,Hsueh_JourMechPhysSol_2018,Jorgensen_IJSS_2017,Mateos_AdvFuncMat_2019,Cui_IJSS_2020,MuroBarrios_JMPS_2022,Fulco_EML_2022,Fulco_JMPS_2024} by altering the stresses around cracks or defects. In particular, lattice structures with fine-scale features are often engineered in such a way that damage at the microscale governs the global failure behavior of the material \cite{Luan_JMPS_2022,OMasta_JMPS_2017,Shaikeea_NatureMaterials_2022,Zhang_NewJourPhys_2018,Athanasiadis_EML_2021,Hedvard_IJSS_2024}. While the failure of simple structures, such as uniform lattices, can be precisely forecast and linked to broader failure characteristics \cite{Fulco_EML_2022,Fulco_JMPS_2024,Hedvard_IJSS_2024,Jorgensen_JMPS_2020}, predicting failure in more complex or disordered structures remains a significant challenge. Disordered materials have shown substantial enhancements in toughness, relative to equivalent ordered materials \cite{Fulco_PNASNexus_2025}, which they achieve through the generation of distributed damage \cite{Driscoll_PNAS_2016}. In these cases global failure analysis is typically performed by measuring the local damage in simulations or experiments~\cite{Fulco_PNASNexus_2025,Berthier_PNAS_2019} or inferred using data-driven methods such as graph neural networks \cite{Karapiperis_ComEng_2023}. However, both of these approaches fail to elucidate the underlying mechanisms of microscale damage and its correlation to global fracture properties.
\par Additionally, the interplay between architecture and non-deterministic material properties must be considered to accurately predict the failure of structured materials. For brittle materials, stochasticity \cite{Weibull_JAM_1951} often plays a significant role in the failure process. This has been shown to result in significant impacts for structured lattice materials, with highly stochastic lattices exhibiting large losses in average strength \cite{Ziemke_MatDes_2024}. For fracture, it has also been shown \cite{Lavoie_MMS_2023} that stochastic failure has a significant effect on the fracture length scale \cite{Irwin_JAM_1957}, $\lambda$, also called the fracto-cohesive length \cite{Chen_EML_2017} of the lattice material. Sufficiently stochastic materials may achieve critical length scales such that cracks that would otherwise cause a fracture-like failure no longer effect the failure loads of the material \cite{Lavoie_MMS_2023}. Thus, in order to fully leverage material architecture to control and enhance the fracture toughness of brittle structured materials, a model is required that integrates the effects of the architected material geometry and stochastic failure.
\par In this work, we present a framework for predicting failure of elastic-brittle materials featuring disordered and irregular lattice-like microstructures. The framework includes considerations for stochastic material failure and demonstrates the connection between geometric variability in the material's structure, and stochasticity in its failure properties. Loads to cause the initiation of damage are predicted rigorously, along with estimates of the fracto-cohesive length and representative volume element size that control whether a fracture-like process occurs. The model is extended to predict crack-tip damage zones that may also be used to estimate the steady-state fracture toughness. Results are verified through finite element simulations of a triangular lattice, which reveal how toughness is enhanced through geometric disorder, while stochastic failure can enhance toughness of ordered lattices but generally reduces the toughness of disordered lattices.
\section{Theoretical Framework}

\subsection{Stresses in a Stretch-Dominated Lattice}
\par A lattice material can be assumed to have a pseudo-continuous global stress state, $\sigma_{ij}$, analogous  to a homogeneous material, provided that the length scale of the lattice's geometric features are much smaller than the specimen size and the characteristic length scale of the stress distribution. Under these conditions, for a 2-D structured material with a lattice microstructure that is stretch-dominated \cite{Fleck_ProcRoySocA_2010}, the axial stress of any isolated ligament oriented at an angle $\phi$ can be calculated as the component of the stress field in that direction. Under the assumption of a stretch-dominated material, this is the only pertinent stress component. Thus, the ligament stress, $\sigma_a$, at a position $(r,\theta)$ is given by
\begin{equation}
    \sigma_a (r,\theta)=R_{1i}R_{1j}\tilde{\sigma}_{ij}(r,\theta), \text{     with     } R_{ij}=
    \begin{Bmatrix}
        \cos{(\phi)} && -\sin{(\phi)}\\
        \sin{(\phi)} && \cos{(\phi)}
    \end{Bmatrix},
\end{equation}

\noindent where $\phi(r,\theta)$ is the angle made by the ligament to the $x$-axis, and $\tilde{\sigma}_{ij}=\bar{\rho}\sigma_{ij}$ is the local stress state resulting from the relative density, $\bar{\rho}$. Given the linear scaling of the local stress with the relative density, it is not difficult to infer behaviors of lattices at varying densities so the model is presented for the general case of $\tilde{\sigma}_{ij}$.
\par
\begin{figure*}[ht]
    \centering
    \includegraphics[width=\textwidth]{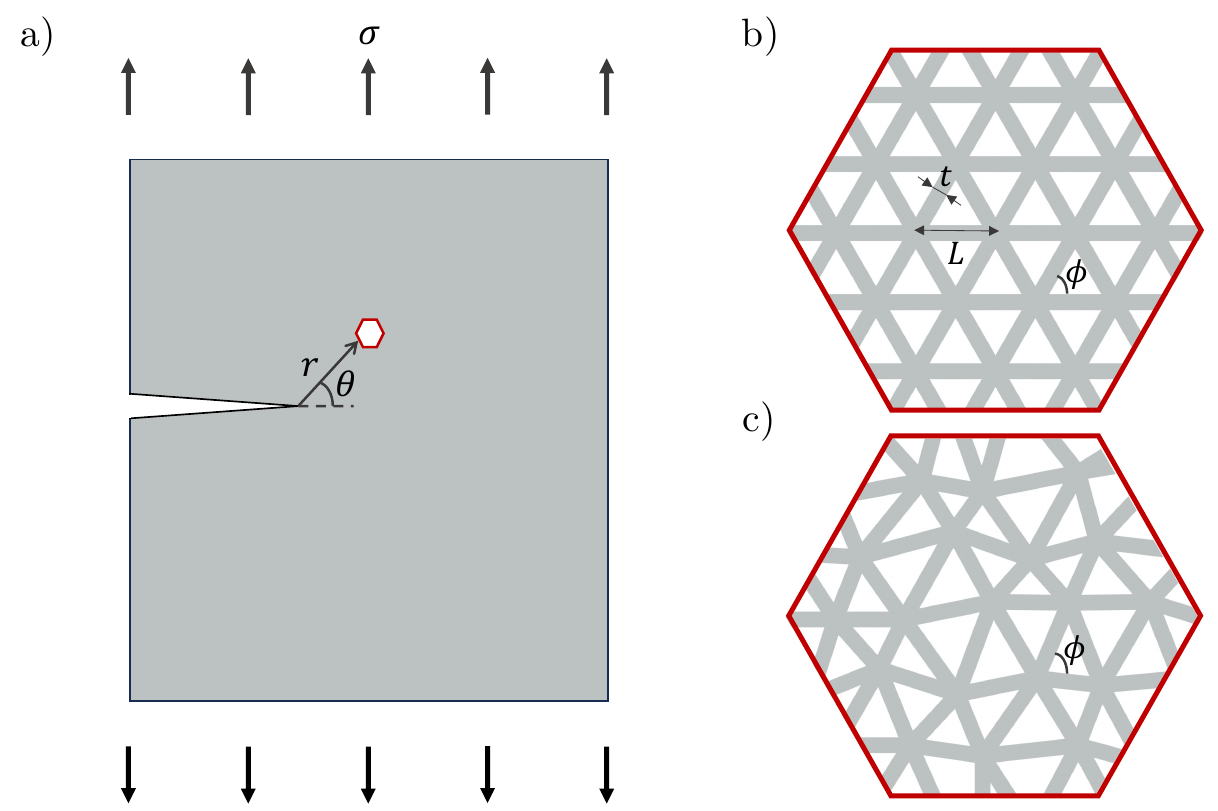}
    \caption{(a) Fracture specimen with crack length $a$, applied far-field stress, $\sigma$, and polar coordinates $r$ and $\theta$ indicated. (b) Ordered triangular lattice corresponding to outlined region. Ligaments have length $L$, thickness, $t$, and are oriented at angles, $\phi$. (c) Disordered triangular lattice}
    \label{fig:fig_1}
\end{figure*}

 As an example, we consider the case of mode I fracture of a linear elastic-brittle material as shown in Fig. \ref{fig:fig_1}(a). In this case, the global stress distribution is singular around a sharp crack tip \cite{Westergaard_JAM_1939}. Under plane stress conditions using polar coordinates centered at the crack tip, these stresses are given \cite{Irwin_JAM_1957} by
\begin{equation}
    \begin{Bmatrix}
    \sigma_{11}\\
    \sigma_{22}\\
    \sigma_{12}\\
    \end{Bmatrix}
    = \frac{K_I}{\sqrt{2\pi r}}\cos{\left(\frac{\theta}{2}\right)}
    \begin{Bmatrix}
        1-\sin{\left(\frac{\theta}{2}\right)}\sin{\left(\frac{3\theta}{2}\right)}\\
        1+\sin{\left(\frac{\theta}{2}\right)}\sin{\left(\frac{3\theta}{2}\right)}\\
        \sin{\left(\frac{\theta}{2}\right)}\cos{\left(\frac{3\theta}{2}\right)}
    \end{Bmatrix},
    \label{eq:Kfield}
\end{equation}
where $K_I=\bar{\sigma}\sqrt{\pi a}$ is the mode I stress intensity factor, with $\bar{\sigma}$ being the far-field applied stress and $a$ the crack length. 
\par The characteristic length scale of the distribution was shown by Irwin  \cite{Irwin_JAM_1957} to be related to the size of the process zone, $\lambda_0$. If the length scale of the lattice is much smaller than the process zone size, $\lambda_0$, it can be assumed that the crack-tip stress field of a homogeneous specimen will be largely maintained at the length scale of the crack and the specimen. Predicting $\lambda_0$ is considered later in this work. Thus, if $\sigma_{ij}$ is given by eq. (\ref{eq:Kfield}), the axial stress in a ligament becomes
\begin{equation}
    \sigma_a=\frac{\tilde{K}_{I}}{2\sqrt{2\pi r}}\cos{\left(\frac{\theta}{2}\right)}\left[\cos{(2\theta-2\phi)}-\cos{(\theta-2\phi)}+2\right],
    \label{eq:sa}
\end{equation}
where $\tilde{K}_I=\tilde{\sigma}\sqrt{\pi a}$. Failure of any individual ligament occurs when its axial stress $\sigma_a=\sigma_f$, where $\sigma_f$ is the failure stress for a brittle material. This is in contrast to an elastic-brittle homogeneous material, where failure via fracture is determined when the stress intensity factor reaches a critical value, $K_{IC}$, the fracture toughness, which is a material property. This is related to the material's toughness \cite{Irwin_JAM_1957} (i.e., the critical energy release rate) by $G_c=K_{IC}^2/\Bar{E}$, where $\Bar{E}=E$ in plane stress and $E/(1-\nu^2)$ in plane strain, where $E$ is the Young's modulus of the material, and $\nu$ is the Poisson's ratio. 
\par In this work, we consider the example case of a triangular 2-D lattice, as shown in Fig. \ref{fig:fig_1}(b), although extensions to other stretch-dominated lattices are also discussed. In a regular triangular lattice with a microscale unit cell size, we can assume that at any location $(r,\theta)$, there will be six ligaments oriented at fixed angles, $\phi=-2\pi/3,-\pi/3, 0, \pi/3, 2\pi/3$, or $\pi$ (assuming no geometric disorder). Thus, ligament failure is deterministically given by the condition $\sigma_a(r,\theta | \phi)\geq \sigma_f$, for any of the six possible values of $\phi$. 
\subsection{Effect of Material Stochasticity}
\par The above analysis is rigorous for an idealized elastic-brittle material; however, many real brittle materials exhibit stochastic failure due to their sensitivity to cracks and defects. This sensitivity is the result of the critical crack size to cause failure by fracture being on the order of the typical defect size in the material. Thus, these materials have three distinct length scales; the specimen and stress distribution scale, the lattice scale, and the microstructural defect scale. For a lattice composed of a brittle material, the resulting variability in failure can significantly affect the distribution of damage during failure, which changes the analysis from a deterministic problem to a probabilistic one. In the case of tensile failure \cite{Ziemke_MatDes_2024}, it has been shown that repeated tests of the same lattice results in damage propagation along varying paths, leading to a range of macroscopic failure behaviors. Stochastic material failure is also shown to significantly reduce the average tensile strength of a lattice \cite{Ziemke_MatDes_2024}.
\par
\begin{figure*}[ht]
    \centering
    \includegraphics[width=\textwidth]{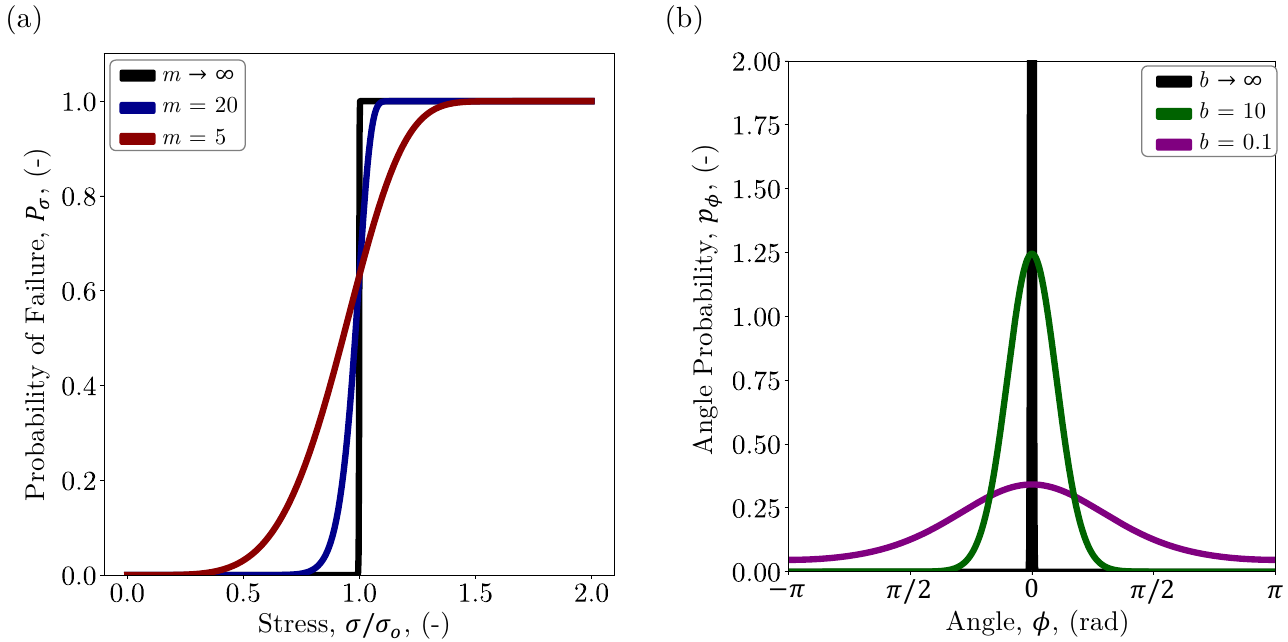}
    \caption{(a) Probability of failure given by Weibull distribution for varying Weibull moduli, $m$, as a function of the ratio of the applied stress to the reference stress, $\sigma/\sigma_o$. (b) von Mises distributions for ligament angles as a function of the disorder parameter, $b$, with $b\rightarrow\infty$ indicating the condition of an ordered lattice.}
    \label{fig:fig_2}
\end{figure*}

 Stochastic failure of materials is typically described by a cumulative distribution function (CDF), such as the Weibull distribution \cite{Weibull_JAM_1951}, which predicts a failure probability, $P_\sigma$, of a material as
\begin{equation}
    P_\sigma(\sigma | m)=1 - e^{-\frac{1}{V_o}\int_V \left(\frac{\sigma}{\sigma_o}\right)^m dV},
    \label{eq:ProbS}
\end{equation}
where $V$ is the volume of the material, $\sigma_o$ is the reference stress measured on a sample of volume $V_o$, and $m$ is the Weibull modulus, which describes the variance in failure behavior. Materials with larger Weibull moduli are less stochastic, as shown in Fig. \ref{fig:fig_2}(a). Many engineering ceramics, for example, have a Weibull modulus near $10$ \cite{Meyers_MechanicalBehaviorOfMaterials_2009}. For simplicity, in this work we will assume that the ligaments are under uniaxial tension and that their volume is comparable to the reference volume of the material, $V\approx V_0$. This simplifies the integral in eq. (\ref{eq:ProbS}), although the framework presented can be easily repurposed without these simplifications. A Weibull distribution is used in this work due to its wide usage for brittle materials, but the analysis is directly adaptable to other distributions suitable for a particular system. The CDF is considered here, rather than the probability density function (PDF), as most real loading conditions will result in a stress increasing from zero to a finite non-zero value, rather than instantaneously achieving the final magnitude. 
\par The CDF provides the probability that failure will occur as the material is loaded to some final stress state. For a lattice material, the probability that a ligament a point $(r,\theta)$ and oriented at an angle $\phi$ will fail at a given load is given by 
\begin{equation}
    P_\sigma(r,\theta | \phi,m)=1-e^{-\left(\frac{\sigma_a(r,\theta|\phi)}{\sigma_o}\right)^m},
    \label{eq:Weibull}
\end{equation}
where the integral in eq. (\ref{eq:ProbS}) was simplified using the assumption of uniaxial stress, and the ligament volume is assumed to be approximately equal to $V_o$. As the failure of the lattice is no longer deterministic, ligament failures cannot be predicted at any location, only the probability that a ligament will have failed can be determined, given the stress in the region and the orientation of the ligament. 
\subsection{Effect of Geometric Variability}
\par The most difficult systems to predict where damage will form are materials with random or disordered structures. Unlike in a regular or periodic system, where a small number of possible features are available and the crack front is self-similar during propagation, a disordered system must consider a variety of possible structures and the probability of their appearance in the material as the crack advances. These details can be encoded in the analysis of the lattice microstructure by considering the ligament angles, $\phi_i$, as a set of random variables described by probability density functions, $f_{\phi_i}$, rather than as a set of deterministic discrete quantities. These probability density functions provide the likelihood of finding a ligament at any particular angle and can be tuned to match the microstructure of the material.
\par For the case of a triangular lattice, disorder can be introduced to the structure by perturbing the node locations. This will result in ligaments oriented at a set of random angles centered around the original discrete angles of the regular lattice but does not change the connectivity, preserving the assumption of stretch-dominated mechanics. The distribution of ligament angles can be described by a von Mises distribution \cite{Fisher_1987}, which is analogous to a normal distribution but is $\pi$-periodic, as shown in Fig. \ref{fig:fig_2}(b). The probability of finding a ligament at an angle $\phi$ is then given by
\begin{equation}
    f(\phi| b,\mu)_{\phi_i}=\frac{1}{2\pi I_0(b)}e^{b\cos{(\phi-\mu)}},
\end{equation}
where $I_0$ is the modified Bessel function of the first kind of order zero, $\mu$ is the mean around which the angle is centered, and $b$ controls the variance, with $b^{-1}$ being qualitatively analogous to the variance in a normal distribution. A geometrically deterministic lattice is therefore given by $b\rightarrow \infty$ (Fig. \ref{fig:fig_2}(b)). This distribution is provided as an example for a randomly disordered triangular lattice, but the approach is adaptable to other distributions that are tailored to specific material systems. 
\par If a node of a lattice has a connected set of ligaments that are likely to be found distributed around a set of $N$ average angles $\mu={\mu_1,\mu_2,...,\mu_N}$, then the probability of finding a ligament at angle $\phi$ is given by the set of probability density functions, $f_{\phi}={f_{\phi_1},f_{\phi_2},...,f_{\phi_N}}$. If ligaments are described by a single distribution, $N=1$, but for the random triangular lattice $N=6$. As the possibility of each of these ligaments failing can be taken as independent of each other, the total probability of ligaments failing can be found by the product of their probabilities of failure. The total probability of failure can then be found by considering the complement of the probabilities of survival. Thus, for a node in the lattice at location $(r,\theta)$, the total probability of failure of any of the adjoining ligaments is then given by 
\begin{equation}
    P_f(r,\theta, \bar\sigma | m, \sigma_o, b)=1 - \prod_{i=1}^6 \left[ 1 - \int_{-\pi}^{\pi}P_\sigma(r,\theta,\phi,\bar\sigma || m,\sigma_o)p_{\phi}(\phi | b,\mu_i)d\phi \right].
    \label{eq:localfailure}
\end{equation}
\par If the lattice is geometrically deterministic (i.e., $b\rightarrow\infty$), equation (\ref{eq:localfailure}) simplifies to  $P_f(r,\theta | m, \sigma_o)=\prod\limits_{i=1}^N P_\sigma(r,\theta| \mu_i, m)$, and only depends on the material stochasticity. If the ordered lattice is deterministic in its failure (i.e., $m\rightarrow\infty$), equation (\ref{eq:localfailure}) further simplifies to a binary, where $P_f(r,\theta)=0$ if $\sigma_a\leq\sigma_0$, or $P_f(r,\theta)=1$ otherwise, which is the process outlined at the beginning of this work for a regular, deterministic lattice.  
\par To quantify the probability that failure has occurred anywhere in the lattice, we treat the failure of the ligaments around any node as distinct from one another and consider the product of their probabilities. For a large lattice, the number of nodes tends towards infinity, suggesting an infinite product. To remain general to varying probability density functions that may not converge for an infinite product, we instead propose a Monte-Carlo approximation of the problem. Considering $s$ nodes located as a set of points $S=\{(r_j,\theta_j)|j=1,...,s, 0 <r_j\leq r_{max}, -\pi\leq\theta\leq\pi\}$, where $r_{max}$ is the furthest radial extent of the specimen, the total probability of failure occurring anywhere in the specimen is given by 
\begin{equation}
    F_f(\bar\sigma | m,\sigma_o, b)\approx 1-\prod_{j=1}^s \left[1-P_f(r_j,\theta_j, \bar\sigma | m, \sigma_o, b)\right].
    \label{eq:totalfailure}
\end{equation}
The above approximation becomes an equality when $s$ accounts for all nodes in the lattice, with $s\rightarrow\infty$ as the lattice becomes large. This relationship provides the CDF that a ligament has failed anywhere in a random, stochastic lattice. However, the expression does not account for changes to the global stress distribution that arises from these failures, and so it can only be rigorously applied to the case of damage initiation, or the failure of one ligament. We demonstrate its predictive capabilities in this capacity through numerical modeling below. When considering a large number of ligament failures, such as if there is significant distributed damage during a fracture process, the model is only approximate, and becomes less accurate for larger number of ligament failures as the stress state will deviate in increasing amounts from the initial assumed state. Instead, we propose an alternative extension of the model to considering distributed damage later in the text. 
\section{Numerical Modeling}
\par
\begin{figure*}[ht]
    \centering
    \includegraphics[width=\textwidth]{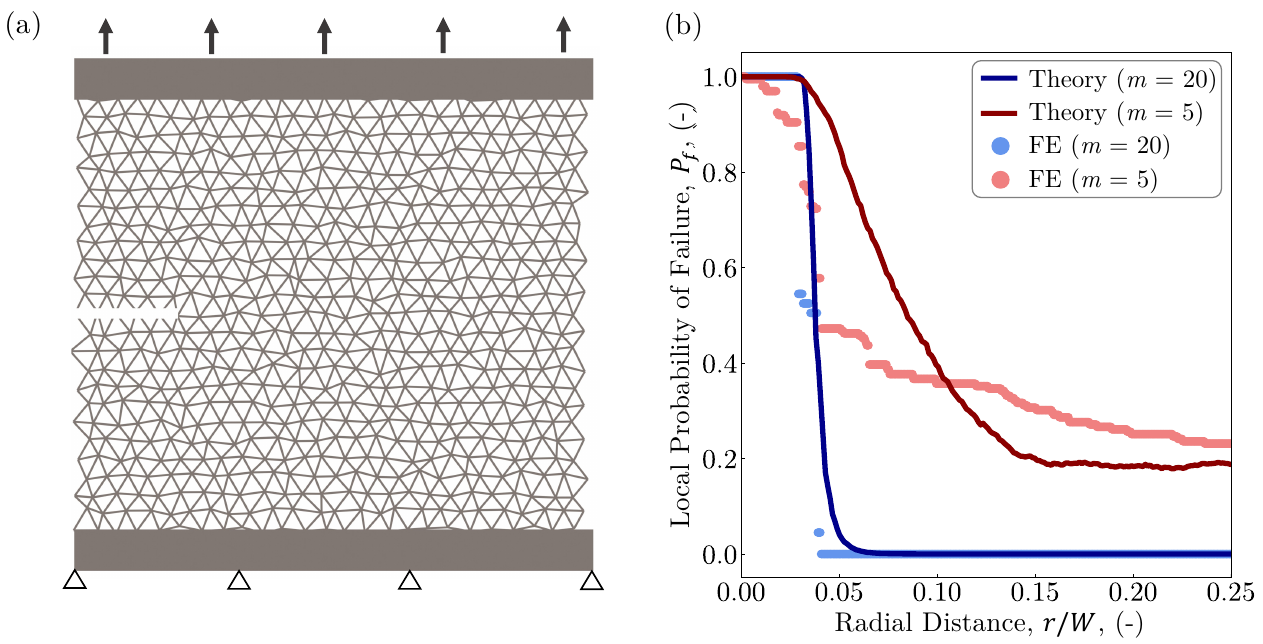}
    \caption{(a) Representative finite element model geometry, consisting of a 30x30 unit cell triangular lattice with a precrack under displacement control boundary conditions. (b) Probability of ligament failure as a function of the radial distance from the crack tip from the finite element (FE) simulations and the analytic moodel.}
    \label{fig:fig_3}
\end{figure*}

 Lattice geometries were generated in the open-source computer-aided design software FreeCAD, implemented through a Python script (v. 3.10). A 30x30 triangular lattice with unit cells of length $L=5$ mm and ligament thickness $t=0.5$ mm were used, resulting in a 150 mm by 130 mm lattice. Two solid 10 mm beams were mounted on the top and bottom to apply a uniform displacement to the lattice, as shown in Fig. \ref{fig:fig_3}(a). To generate disordered lattices, nodes were perturbed at random angles with a controlled distribution of radial magnitudes to match the case of a von Mises distribution with $b=10$, as shown in Fig. \ref{fig:fig_3}(a). Models were then exported as 2-D DXF drawing files and imported into the finite element software.
\par Finite element simulations were performed in ABAQUS (2020, Providence, RI) using 2-D plane stress conditions (specimen thickness was taken as 6 mm), with the lower mounting point in the solid beams being fixed in $x$ and $y$, and one under displacement boundary conditions in $y$ and fixed in $x$. The models were meshed with bilinear plane-stress quadrilateral elements (CPS4), with a free mesh in the lattice and a uniform mesh in the beams, with a mesh size of 0.05 mm. Approximately 160,000 elements were used per simulation, depending on the exact geometry. 
\par The material was modeled with an elastic modulus of $E=1$ GPa, and the reference stress was taken as $\sigma_o=20$ MPa. A 1 mm displacement was applied to the upper solid beam. The average von Mises stress over an area with a radius of $2t$ was calculated at the center of every ligament. The probability of survival was calculated for every ligament using eq. (\ref{eq:ProbS}) and compared to a set of randomly selected real numbers between 0 and 1 using SciPy's inbuilt random number generator. If any of the ligaments had a probability of survival lower than their associated random number, they are identified as a failed ligament. If multiple ligaments fail, then the ligament with the lowest probability of survival was taken as the broken ligament. If no ligaments fail, then the displacement is incremented by 0.5 mm, and a new probability of survival for each ligament is calculated. This process is iterated until a ligament fails. One hundred simulations were performed for the cases of an ordered lattice with $m=20$, an ordered lattice with $m=5$, and a disordered lattice ($b=10$) with $m=5$, for a total of three hundred simulations. 
\par Validation of the finite element simulations and the analytic model was performed by comparing the prevalence of ligament failures in the simulations as a function of radial distance to that predicted by eq. (\ref{eq:localfailure}). To model the failure throughout the entire lattice structure, a piecewise global stress distribution appropriate for an edge-cracked specimen was used. In this stress distribution, the \emph{K}-field stresses given by eq. (\ref{eq:Kfield}) are used near the crack tip which then decays to the the far-field applied stress, which is then taken as a fixed value determined by the boundary conditions.
\par The probability of ligament failure of the analytic model and the finite element simulations are shown in Fig. \ref{fig:fig_3}, as a function of radial distance from the crack tip. Results are given for two example cases of ordered lattices with two different Weibull moduli, $m=20$ and $m=5$. The simulations and analytic model show good agreement and demonstrate how damage is localized near the crack tip for the more deterministic case of a high Weibull modulus, and becomes more distributed for a low Weibull modulus. This is expected since the high stresses near the crack tip should result in a large probability of failure, but the lower stresses farther from the crack tip only result in a meaningful number of failures for the more stochastic material. There is some small disagreement between the model and simulations for the more stochastic case at short distances from the crack tip. This is expected due to the assumption of a singular stress state that results in a probability of failure approaching 1 for any Weibull modulus. As the finite element model with a crack tip of finite radius does not achieve a true singular stress, it is expected that the model should under-predict the probability of failure near the crack tip, relative to the singular analytic model, as is observed.

\section{Results: Stochasticity, Disorder, and Length Scales}

\subsection{Failure Loads}
\begin{figure*}[ht]
    \centering
    \includegraphics[width=\textwidth]{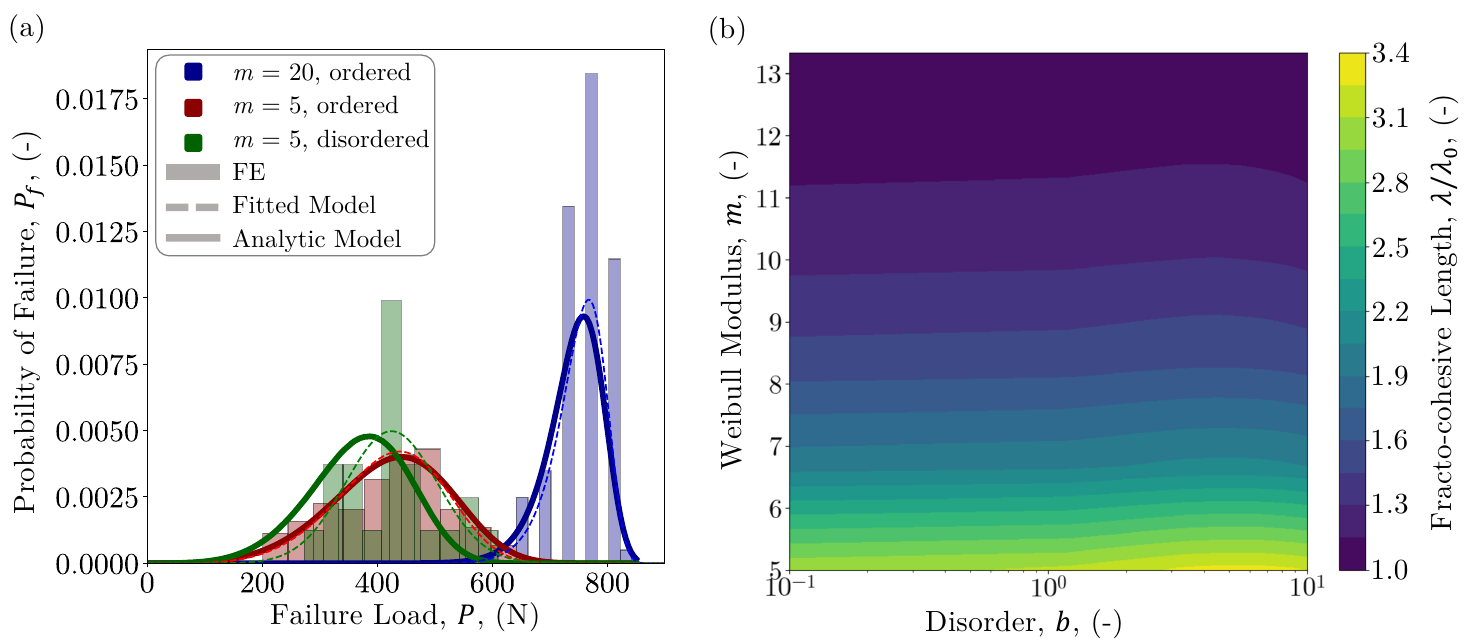}
    \caption{(a) Probability distribution of ligament failures in a lattice as a function of the applied load. Histogram results given for finite element simulations, along with analytic model predictions and fitted models for each dataset (Weibull for ordered datasets, Gaussian for disordered dataset). (b) Fracto-cohesive length relative to the length for an ordered, deterministic lattice, as a function of Weibull modulus, $m$, and disorder, $b$.}
    \label{fig:fig_4}
\end{figure*}

The model in eq. (\ref{eq:totalfailure}) is rigorous for the initiation of damage (i.e., the first ligament failure of the lattice). Figure \ref{fig:fig_4}(a) shows the prevalence of failure loads measured from the finite element simulations for ordered lattices with Weibull moduli $m=20$ (pseudo-deterministic), and $m=5$ (stochastic), as well as a disordered lattice also with Weibull modulus $m=5$. As shown in Fig. \ref{fig:fig_4}(a),  stochasticity has a significant impact on the average failure load and its variation, with the $m=20$ ordered lattice having a failure load of $750 \pm 43$ N (mean $\pm$ std. dev.), while the ordered lattice with $m=5$ has a failure load of $420\pm 95$ N. Note that the relative magnitude of the standard deviation to the mean failure load increases by nearly a factor of four between the two cases. The introduction of disorder to the $m=5$ case results in no significant change in failure load or variation, with an average failure load of $425\pm 80$ N.
\par By taking a discrete derivative of eq. (\ref{eq:totalfailure}) with respect to the far-field stress, we can calculate the PDF of the failure loads. These predictions are given in Figure \ref{fig:fig_4}(a) and show excellent agreement with the simulation results in all cases, accurately capturing both the mean and variance of the failure loads. Also shown are best-fit models to the data, considering either a Weibull distribution or normal distribution, as was done in \cite{Ziemke_MatDes_2024}. A maximum log-likelihood criterion was used to determine the quality of the fit. For the ordered lattices the best-fit model is the Weibull distributions with fitted moduli values of 20.0 and 5.0, exactly. This is expected from eq. (\ref{eq:totalfailure}), as it is simple to see that in the limit of an ordered lattice the probability density functions for $\phi$ collapse to delta functions, resulting in a simple product of Weibull distributions. For the disordered lattice, the normal distribution achieved a better quality of fit, as was also observed in \cite{Ziemke_MatDes_2024}. This is now explained by eq. (\ref{eq:totalfailure}), which shows that, while the data is not expected to truly follow a normal distribution, it will also not follow a pure Weibull distribution as the stochastic behavior is modulated by the geometric disorder. 
\subsection{Fracto-cohesive Length}
The results in Fig. \ref{fig:fig_4}(a) demonstrate that the analytic model accurately predicts the required load to initiate damage in a lattice material. This result can be utilized to calculate additional parameters that characterize the failure of the lattice, such as the fracto-cohesive length, $\lambda$. If any defect in the lattice is larger than $\lambda$, the material will undergo a fracture-based failure, rather than a strength-based failure \cite{Chen_EML_2017}. Knowing which regime a system is in can be useful for making predictions about failure. For example, if a lattice will undergo a fracture-based failure, the failure loads will strongly depend on the defect size, unlike in the strength-based failure regime.
\par The fracto-cohesive length is defined \cite{Chen_EML_2017} as 
\begin{equation}
    \lambda = \frac{G_c}{W_f}
    \label{eq:lambda}
\end{equation}
where $W_f=\int_{0}^{\epsilon_c}\sigma(\epsilon)d\epsilon$ is the work of fracture for an uncracked material, where $\epsilon_f$ is the failure strain. For a linear elastic material that fails in a brittle manner under uniaxial tension, this expression simplifies to $W_f=\frac{\sigma_f^{\ast^2}}{2E}$, where $\sigma_f^\ast$ is the failure stress of the lattice material. Note that this is distinct from the failure stress of the base material of the lattice, $\sigma_f$. This can be calculated directly from eq. (\ref{eq:totalfailure}), where the local stress field is taken as a constant ($\sigma(r,\theta)=\bar\sigma$). 
\par The toughness, $G_c$, can be estimated from eq. (\ref{eq:totalfailure}) as well using the Irwin-Kies relationship, with $G_c=K_{IC}^2/E$, in plane stress, where $K_{IC}=Y\sigma_c^\ast\sqrt{\pi a}$ is the critical stress intensity factor, with $Y\approx1$ is a geometry factor, and $\sigma_c^\ast$ is the critical stress to cause fracture for the given crack length, which can be predicted from eq. (\ref{eq:totalfailure}). Therefore, the fracto-cohesive length can be estimated as
\begin{equation}
    \lambda = \left(\frac{K_{IC}}{\sigma_f^\ast}\right)^2. 
\end{equation}
\par Figure \ref{fig:fig_4}(b) shows predictions of the fracto-cohesive length from eq. (\ref{eq:lambda}), where $G_c$ is calculated from eq. (\ref{eq:totalfailure}) using the \emph{K}-field stress distribution, and $W_c$ is calculated using a uniform tensile stress. Results are normalized by the predicted fracto-cohesive length for an ordered, deterministic lattice of equivalent density, $\lambda_0$. The results show that the fracto-cohesive length depends strongly on the Weibull modulus, as shown in \cite{Lavoie_MMS_2023}, but that had not been quantitatively explained before. Further, there is a dependence on the geometric disorder, although it's effect is far smaller. This makes sense since the stresses can only be affected marginally by the angles of the ligaments if they are randomly oriented, whereas the changes in stochasticity can significantly impact the probability of failure given a local stress.
\subsection{Representative Volume Element}
\par Beyond the fracto-cohesive length, the size of a full representative volume element (RVE) for a lattice can be calculated. This has a theoretical utility as a test of whether a lattice is sufficiently large for the model to be accurate. If a lattice is much smaller than its RVE, the assumption of a pseudo-continuous stress field (eq. (\ref{eq:sa})) will not apply. A practical utility of the calculation is to determine if a lattice is much larger than the RVE, as its mechanics can then be homogenized, which is useful for characterizing the global behavior of the material, as shown below. 
\par Considering a simple rectangular RVE that spans the width of the material, $b$, with an area $h\xi$, where $\xi$ is the length of the RVE along the crack path ($\theta=0$) and $h$ is the height of the RVE perpendicular to the crack path ($\theta=\pi/2$). In a strict sense, the length of the RVE, $\xi$, needs to only span the distance over which the crack advances. In the case of an ordered, deterministic lattice material, the crack will jump in discrete intervals equal to the length of a unit cell, $L$. For a lattice material with disorder, this may not be the case and damage may occur in a unit cell not directly in front of the crack tip. By defining the RVE length using the first-order Irwin fracture stress length scale, $\lambda_0=(2\pi)^{-1}(K_{IC}/\sigma_f)^2$, ensures that the entire region of high stress around the crack tip is encompassed where most damage is expected to occur. Inspired by the approach of Irwin, this length scale can be adapted to a stochastic material by considering a length ahead of the crack tip, $\xi$, where the probability of failure is $T$ (taken to be sufficiently close to $1$). From eq. (\ref{eq:Weibull}) for $\theta=0$, it is easy to show that 
\begin{equation}
    \xi =  \frac{1}{2\pi}\left(-\ln{(1-T)}\right)^{-\frac{2}{m}}\left(\frac{K_{IC}}{\sigma_0}\right)^2.
\end{equation}
It is clear that as $m\rightarrow\infty$, $\xi/\lambda_0 \rightarrow 1$, as expected. The choice of $T$ is arbitrary, and so this length scale should not be treated as an inherent material property like $\lambda$; however, for reasonable values of $m$ ($m>1$), choices of $T$ between 90\% and 99.99\% are of the same order of magnitude and can therefore  provide an estimate of the RVE. In this work we will consider the conservative case of $T=0.9$, which will create a larger RVE.
\par The height of the RVE defines the region above the crack in which the majority of the energy is released during fracture. Following the calculation in \cite{Purohit_JMPS_2025}, the height of the RVE can be found as $h=G_c/W_c$, where $W_c=\frac{1}{2}\sigma_{ij}\epsilon_{ij}$ is the critical strain energy density in the RVE at the point of crack propagation. Using the \emph{K}-field stresses given in eq. (\ref{eq:Kfield}), and the corresponding strains for the case of plane strain, the critical strain energy density in front of the crack ($\theta=0$), becomes
\begin{equation}
    W_c=\frac{K_{IC}^2}{E^\ast}\left(\frac{1-\nu^\ast}{2\pi \xi}\right),
    \label{eq:Wc_initial}
\end{equation}
where $E^\ast$ and $\nu^\ast$ are the effective modulus and Poisson's ratio of the lattice material. For simplicity, we consider the case of a geometrically ordered, stochastic lattice, and estimate the fracture toughness, $K_{IC}$, using the stress at which a ligament fails at $r=\xi$. (Note, a more thorough estimate of $K_{IC}$ is presented below.) In an ordered lattice, at the point $r=\xi$ and $\theta=0$, the first ligament to fail is the one oriented at $\phi=\pi/3$, resulting in $K_{IC}\approx \sigma_f \sqrt{2\pi \xi}$. Substitution of this result into eq. (\ref{eq:Wc_initial}) leads to 
\begin{equation}
    W_c=\frac{\sigma_f^2}{E^\ast}(1-\nu^\ast).
    \label{eq:Wc_final}
\end{equation}
Using the Irwin-Kies relationship to define the lattice toughness as $G_c=K_{IC}^2/E^\ast$, the height of the RVE can be found as
\begin{equation}
    h=\left(\frac{2\pi}{1-\nu^\ast}\right)\xi.
\end{equation}
For the case of a triangular lattice, $\nu^\ast\approx1/3$ and $h=3\pi\xi$, or the aspect ratio of the RVE is $h/\xi = 3\pi$. For triangular lattices of with unit cells of width $L$, the height of a unit cell is $\sqrt{3}L/2$. Thus, if the RVE width is $\xi=N$ unit cells, then the height is $h=2\sqrt{3}\pi N\approx 11N$. For the lattices considered in our finite element simulations, the vast majority of ligament failures are localized to the first $2.5 \pm 0.5$ unit cells ahead of the crack tip, as shown in Fig. \ref{fig:fig_3}, which suggests that the RVE height must be $27\pm 5$ unit cells high.
\begin{figure*}[ht]
    \centering
    \includegraphics[width=\textwidth]{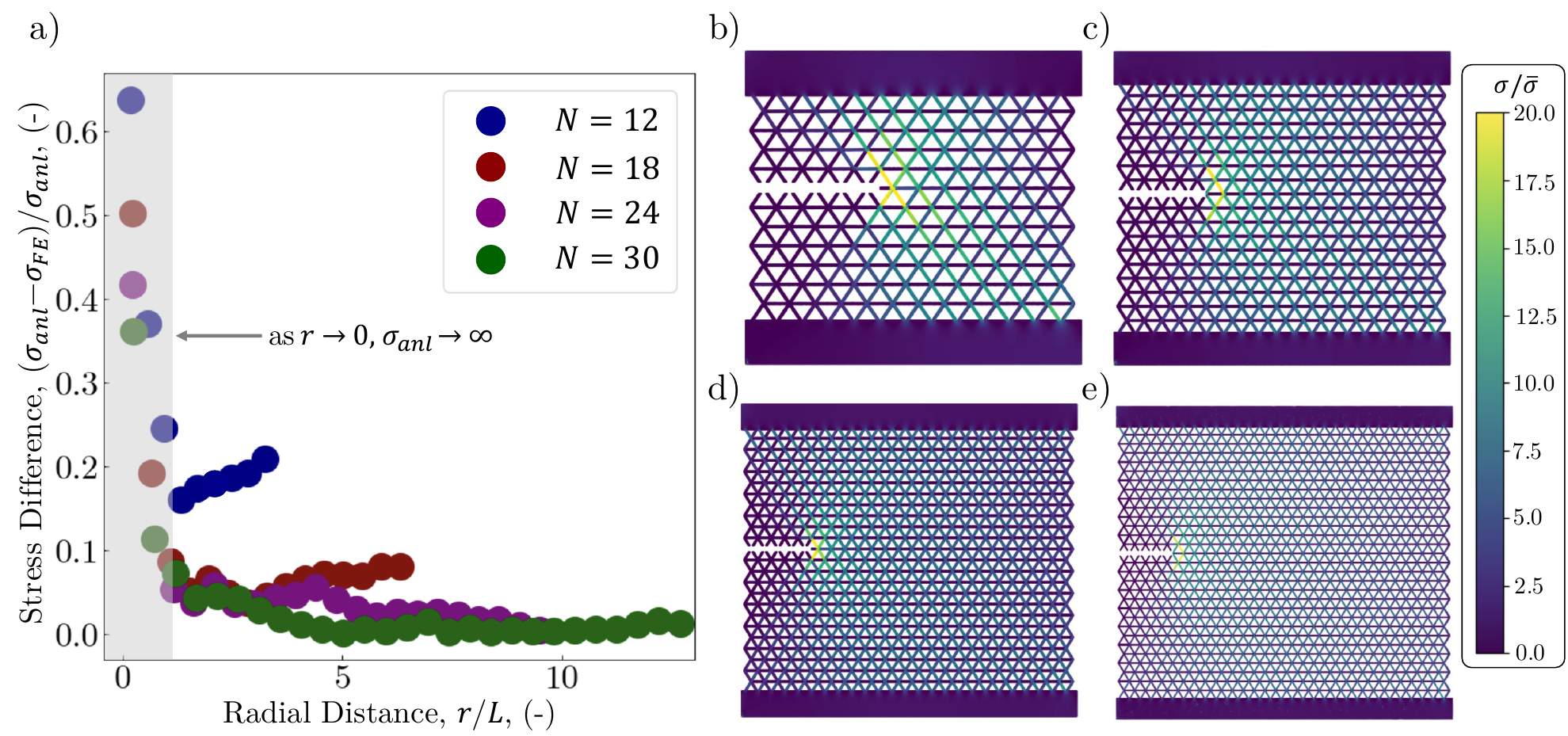}
    \caption{a) Average difference in ligament axial stresses between the analytic model ($\sigma_{anl}$) and the finite element model ($\sigma_{FE}$) as a function of the radial distance from the crack tip for four lattice sizes, $N=12, 18, 24,$ and $30$, shown in (b-e), respectively.}
    \label{fig:fig_6}
\end{figure*}
\par Lattices of varying numbers of unit cells were simulated to determine what the minimum RVE size should be and to verify the prediction of $h$ from the model. Ordered lattices with deterministic failure were considered. Minimum RVE size can be determined independently by measuring the local ligament stresses and comparing to the predictions of a \emph{K}-field stress. Fig. \ref{fig:fig_6}(a) shows that the average difference in ligament axial stresses between the finite element model and the analytic predictions from eq. (\ref{eq:sa}) as a function of the radial distance from the crack tip. Results are given for four lattices of varying size, with $N=12, 18, 24$ and $30$, as shown in Fig. \ref{fig:fig_6}(b-e). For all of the lattices there is disagreement between the analytic model and the finite element model for radial distances smaller than one unit cell in length as $r\rightarrow 0$. This is expected since the idealized \emph{K}-field stresses become infinite, which is not physical. However, at greater radial distances it is clear that the large lattices agree more accurately with the analytic model. The $N=12$ lattice maintains an approximately 20\% deviation from the analytic model across the entire lattice, and the $N=18$ lattice has a nearly 10\% error across most of the lattice. It is only for $N=24$ that the error drops below 5\% and for $N=30$ the error converges to $<1\%$ for larger radial distances. This is in line with the predictions of $h$ required to obtain the asymptotic crack tip field in an architected material and is an independent verification of the model prediction. 
\section{Results: Toughness and Damage Zones}
\par For the case of a lattice that is much larger than its RVE, its mechanics can be homogenized, which allows the model to be extended beyond predicting failure loads to estimating fracture toughness. As noted above, eq. (\ref{eq:totalfailure}) is rigorous in predicting failure for the initiation of damage in a lattice structure. However, it is apparent from the results in Fig. \ref{fig:fig_3}(b) that significant distributed damage can occur in these lattices, suggesting that the steady-state toughness of the material would likely not be achieved until a large number of ligament failures occur, such that the \emph{K}-field stresses shift along the crack path. As more ligament failures occur, eq. (\ref{eq:totalfailure}) will become less and less accurate as it does not account for the changing stress state due to the change in local connectivity throughout the lattice. However, in the case of a large lattice where the mechanics are homogenized, the model can instead be extended to predict the toughness by using the distribution of damage, as has been shown previously to be an accurate predictor of the effective toughness of the material \cite{Fulco_JMPS_2024,Fulco_PNASNexus_2025}. 
\par While in a stochastic lattice damage may theoretically occur anywhere, the singular nature of the \emph{K}-field stresses ensures that the majority of damage occurs near the crack-tip (see Fig. \ref{fig:fig_3}(b)). Since the stress fields are monotonically decreasing as the radial distance increases, it enables us to define a region within which there is a threshold level of probability of failure, $T$, outside of which the probability of failure is lower. This is then analogous to a process zone in a homogeneous material.
\par
\begin{figure*}[ht]
    \centering
    \includegraphics[width=\textwidth]{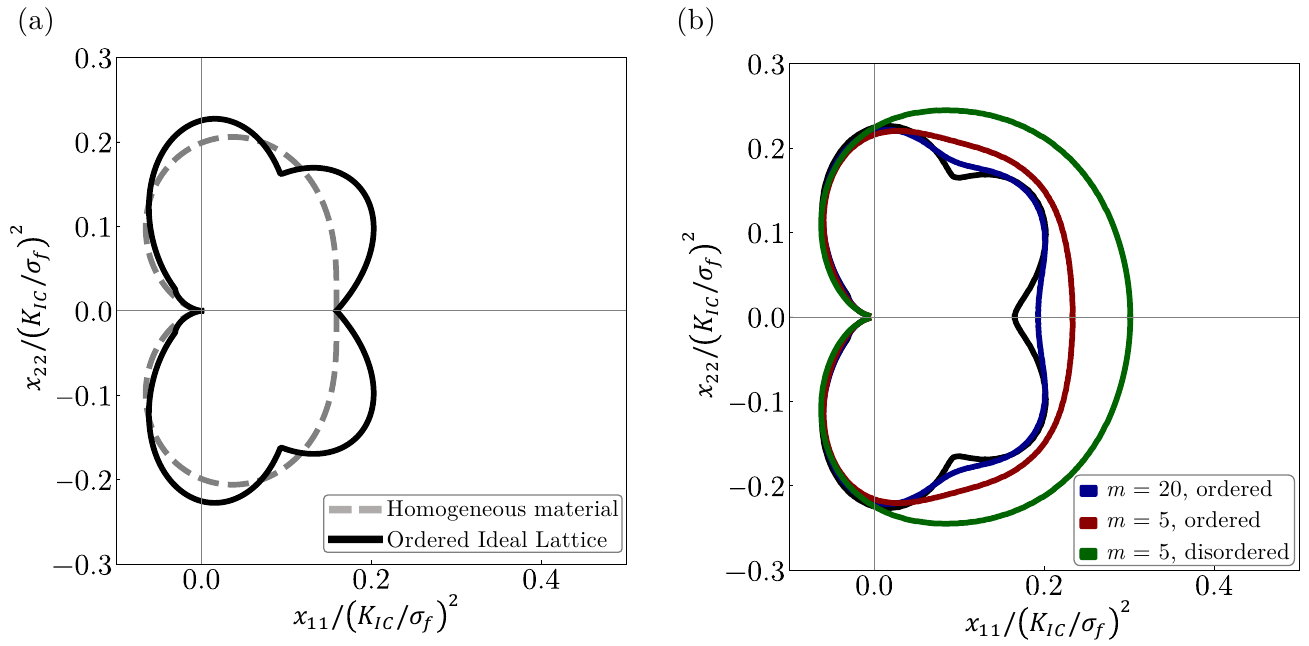}
    \caption{(a) Damage zone for a homogeneous plane-stress material and an elastic-brittle ordered triangular lattice. (b) Damage zones corresponding to a 90\% chance of failure for stochastic lattices with Weibull modulus, $m$, and ordered vs. disordered lattice geometries.}
    \label{fig:fig_5}
\end{figure*}

 The probability of failure is calculated using eq. (\ref{eq:localfailure}) as the radial distance from the crack tip increases and for all $\theta$. A threshold probability of failure, $T$, is chosen to define the damage zone. In this work a probability of failure $ T\geq0.9$ is used. The choice of the threshold is arbitrary, although we note that there was not a significant change in the damage zone sizes as the threshold was increased due to the singular nature of the stress fields. As a control study, the plane stress deterministic damage zones for a homogeneous material, and an ordered lattice are shown in Fig. \ref{fig:fig_5}(a). The lattice damage zone follows closely with the homogeneous material, with perturbations due to  geometric coupling with the six individual ligament angles. 
\par Figure \ref{fig:fig_5}(b) shows the process zones inside of which the probability of failure $F\geq 0.9$ for the cases given in Fig. \ref{fig:fig_4}(a) ($m=20$ and $m=5$ ordered lattices and a $m=5$ disordered lattice, $b=10$). It is apparent that as the network becomes more geometrically disordered, the process zone continues to increase in size as a result of the variability in the ligament stresses due to the geometric disorder, and the variability in failure stresses that result from the stochastic failure. It is also apparent how the introduction of geometric disorder and stochastic failure results in a more isotropic process zone, with the angular dependence seen in the regular lattice replaced by a more uniform zone.
\subsection{Estimating Material Toughness}
\par As shown in Fig. \ref{fig:fig_5}, increasing geometric variability and material stochasticity both tend to increase the size of the process zone in a lattice material. Larger process zones result in an increase in the number of ligament failures, increasing the energy dissipated during crack propagation. As shown in \cite{Fulco_EML_2022,Fulco_JMPS_2024}, the toughness depends on the size of the process zone and maximum local strain energy density, with  $G_c \propto \rho_{e} \Omega$, where $\rho_{e}$ is the energy density of the process zone during failure, and $\Omega$ is the volume of the process zone. 
\par For a linear elastic fiber or lattice material, the maximum energy density in the process zone depends on the local stress before failure. If failure is stochastic and determined by a probability, $P_s$,  then the average ligament energy density will be $\rho_e=\mathbb{E}_{\sigma^2}$, where $\mathbb{E}_{\sigma^2}=\int_{0}^{\infty}P_s'(\sigma|m)\sigma^2 d\sigma$ is the expectation value of $\sigma^2$, where $P_s'(\sigma|m)=\frac{d}{d\sigma}P_s(\sigma|m)$. Thus, the toughness of a stochastic disordered lattice, relative to its deterministic regular analog is
\begin{equation}
    \frac{G_c}{G_o}=\frac{\mathbb{E}_{\sigma^2}}{\sigma_o^2}\left(\frac{\Omega}{\Omega_o}\right),
    \label{eq:GGoSto}
\end{equation}
where $G_o$ and $\Omega_o$ are the toughness and process zone size of the deterministic lattice, respectively. If $P_s$ is the Weibull distribution, the expectation value will be $\mathbb{E}_{\sigma^2}=\sigma_o^2\Gamma\left(\frac{2+m}{m}\right)$, where $\Gamma$ is the Gamma function, which will be $\leq 1$ for all positive values of $m$. Thus, $\mathbb{E}_{\sigma^2}/\sigma_o^2 \leq 1$ and the introduction of material stochasticity will reduce the local strain energy density, which in turn will have the tendency to lower the toughness. However, as shown in Fig. \ref{fig:fig_5}(b), the process zone volume increases with stochasticity, tending to increase the toughness. Introducing geometric variability will also increase the process zone but has a much smaller effect on the local energy density. Thus, geometric variability is expected to generally increase the toughness, while stochasticity introduces competing effects. Specifically for the Weibull distribution in a triangular lattice, the loss in local strain energy will quickly become comparable to the increase in process zone size as $m$ decreases, suggesting limited enhancements in toughness as a result of the loss of strain energy, even as the damage zones are enlarged.

\section{Conclusions}
The failure of lattice metamaterials with geometric disorder and stochastic material failure is predicted from an analytic model that reveals the effects of local geometry and material properties on the mean and variance in the failure loads of the material, as well as the fracto-cohesive length and the minimum RVE size that control whether a fracture-like failure occurs in the material. Increasing stochasticity significantly reduces the failure loads and results in large fracto-cohesive lengths controlling whether failure depends on initial defect size. Geometric disorder has a smaller effect on failure but changes the variation in failure loads from being a pure Weibull-like distribution to a more complicated model better approximated by a normal distribution. An extension to a full fracture model is presented that uses predictions of the damage zones around the crack tip to estimate the toughness of the lattice. Both stochasticity and disorder result in larger damage zones around the crack tip, but stochasticity significantly reduces the local strain energy density, which reduces the toughness and suggests that while disorder does not significantly influence the initiation of failure, it provides an avenue for enhancing the steady state toughness of the lattice. 
\par These results demonstrate how the failure of lattice mechanical metamaterials is influenced by the interplay of the base material failure properties, such as stochastic failure, and the microstructural geometry. Results are presented for a precracked triangular lattice with Weibull stochastic failure, but the framework is easily adaptable to varying lattice geometries, failure criteria, and stress distributions. An analytical model provides predictions of the failure stress and effective toughness of a lattice, as well as details about the nature of the failure (i.e., strength-based or fracture-based). Unlike with data-driven methods, the analytical approach here provides insights into how the global mechanical behavior of metamaterials arises from the combination of material properties and microstructural geometry, along with the quantitative predictions of the failure. 

\section{Acknowledgments}
This research was funded primarily by the National Science
Foundation (NSF) MRSEC program [awards DMR-2309043 and
DMR-1720530] to K.T.T and P.K.P. S.F. acknowledges support from the US
Department of Defense (DoD) through the National Defense
Science \& Engineering Graduate (NDSEG) Fellowship Program.

\bibliographystyle{elsarticle-num}\biboptions{compress}
\bibliography{Biblio}

\end{document}